\documentclass[aps,pre,superscriptaddress,longbibliography]{revtex4-1}
\usepackage{graphicx}
\usepackage{dcolumn}
\usepackage{amssymb,amsfonts,amsmath}

\begin{document}

\title{Fluctuations in protein aggregation: Design of preclinical screening for early diagnosis of neurodegenerative disease} 
\author{Giulio Costantini} 
\affiliation{Center for Complexity and Biosystems,
Department of Physics, University of Milano, via Celoria 26, 20133 Milano, Italy}
\author{Zoe Budrikis}
\affiliation{Institute for Scientific Interchange Foundation, Via Alassio 11/C, 10126 Torino}
\author{Alessandro Taloni} 
\affiliation{Center for Complexity and Biosystems,
Department of Physics, University of Milano, via Celoria 26, 20133 Milano, Italy}
\author{Alexander K. Buell}
\affiliation{Institute of Physical Biology, University of D\"usseldorf, Universit\"atsstr. 1, 40225 D\"usseldorf}
\author{Stefano Zapperi} 
\affiliation{Center for Complexity and Biosystems,
Department of Physics, University of Milano, via Celoria 26, 20133 Milano, Italy}
\affiliation{Institute for Scientific Interchange Foundation, Via Alassio 11/C, 10126 Torino}
\affiliation{CNR - Consiglio Nazionale delle Ricerche,  Istituto di Chimica della Materia Condensata e di Tecnologie per l'Energia, Via R. Cozzi 53, 20125 Milano, Italy}
\affiliation{Department of Applied Physics, 
Aalto University, P.O. Box 11100, FIN-00076 Aalto, Espoo, Finland}
\author{Caterina A. M. La Porta}\email{caterina.laporta@unimi.it}\affiliation{Center for Complexity and Biosystems,
Department of Biosciences,  University of Milano, via Celoria 26, 20133 Milano, Italy}



\begin{abstract}
Autocatalytic fibril nucleation has recently been proposed to be a determining factor for the spread of
neurodegenerative diseases, but the same process could also be exploited to amplify minute quantities of 
protein aggregates in a diagnostic context. Recent advances in microfluidic technology allow analysis of protein aggregation in micron-scale samples potentially enabling such diagnostic approaches, but the theoretical foundations for the analysis and interpretation of such data are so far lacking. Here we study computationally the onset of protein aggregation in small volumes and show that the process is ruled by intrinsic fluctuations whose volume dependent distribution we 
also estimate theoretically. Based on these results, we develop a strategy to quantify in silico the statistical errors associated with the detection of aggregate containing samples. Our work opens a new perspective on the forecasting of protein 
aggregation in asymptomatic subjects.
\end{abstract}
\flushbottom
\maketitle
%
%
\thispagestyle{empty}


\section{Introduction}

The presence of aberrant conformations of the amyloid $\beta$ peptide and the protein $\alpha$-synuclein is considered to be a key factor behind the development of Alzheimer's and Parkinson's diseases, respectively. The polymerization kinetics of these proteins has been shown to consist of nucleation and growth processes and to be strongly accelerated by the presence in 
solution of pre-existing fibrils~\cite{Jarrett1993, Buell2014}, thereby circumventing the slow primary nucleation of aggregates. It was found that surfaces, 
such as lipid bilayers~\cite{Grey2015, Galvagnion2015} and hydrophobic nanoparticles~\cite{Vacha2014} can accelerate the nucleation process dramatically. 
Indeed, in the case of $\alpha$-synuclein, it was found that in the absence of suitable surfaces, the primary nucleation rate is undetectably slow~\cite{Buell2014}. 
Under certain conditions, the surfaces of the aggregates themselves appear to be able to catalyse the formation of new fibrils, leading to autocatalytic behavior and 
exponential proliferation of the number of aggregates~\cite{Ruschak2007, Cohen2013, Buell2014}. This so-called secondary nucleation process is likely to play an important 
role in the spreading of aggregate pathology in affected brains~\cite{Peelaerts2015}, as the transmission of a single aggregate into a healthy cell with a pool of soluble 
protein might be sufficient for the complete conversion of the soluble protein into aggregates.   

An intriguing idea is to exploit this observation to screen biological samples based on the presence of very low concentrations of aggregates for pre-clinical 
diagnosis of neurodegenerative diseases. Indeed, this has been achieved in the case of the prion diseases in a methodology called prion misfolding cyclic 
amplification~\cite{Morales2012}, which is based on the amplification of aggregates through repeated cycles of mechanically induced fragmentation and growth. 
Recently, the applicability of this approach to the detection of aggregates formed from the amyloid $\beta$ peptide has been demonstrated~\cite{Salvadores2014}. 
Furthermore, the autocatalytic secondary nucleation of amyloid $\beta$ fibrils has been exploited to demonstrate the presence of aggregates during the lag phase 
of aggregation~\cite{Arosio2014}.

However, none of these methods currently allow to easily determine the absolute number of aggregates in a given sample. One strategy to address this problem is 
to divide a given sample into a large number of sub volumes and determine for each of the sub-volumes whether it contains an aggregate or not. Due to advances 
in microfluidic technology and microdroplet fabrication~\cite{Theberge2010}, it is now possible to monitor protein aggregation in micron-scale samples~\cite{Knowles2011}, 
a technique that could be used to design microarrays targeted for protein polymerization assays. To be successful this program needs guidance from theory 
to quantify possible measuring errors due to false positive and negative detection. Current understanding of protein polymerization is based on mean-field 
reaction kinetics that have proved successful to describe key features of the aggregation process in macroscopic samples \cite{Knowles2009,Cohen2011,Cohen2013}. 
This theory is, however, designed to treat the system in the infinite volume limit, where the intrinsic stochasticity of the nucleation processes cannot manifest 
itself, so that its applicability to small volume samples is questionable. The importance of noise in protein aggregation was clearly illustrated 
in Ref. \cite{Szavits-Nossan2014}, who proposed and solved the master equation kinetics of a model for polymer elongation and fragmentation, obtaining good 
agreement with experiments on insulin aggregation
~\cite{Knowles2011}. 
 
Here we address the problem by numerical simulations of a three dimensional model of diffusion-limited aggregation of linear polymers \cite{budrikis2014}, including explicitly auto-catalytic secondary nucleation 
processes \cite{Ruschak2007, Cohen2013, Buell2014}. A three dimensional model overcomes the limitations posed
by both mean-field kinetics \cite{Knowles2009,Cohen2011,Cohen2013} and master equation approaches \cite{Szavits-Nossan2014},
which do not consider diffusion and spatial fluctuations. Most practical realizations of protein aggregation reactions are not diffusion limited, due to the slow nature of the aggregation steps, caused by significant free energy barriers \cite{Buell2012}. This leads to the system being well mixed at all times and mean-field theories
providing a good description. There are, however, cases both {\it in vitro} (e. g. when protein
concentrations and ionic strengths are high, leading to gel formation \cite{Buell2014}) and {\it in vivo}
(due to the highly crowded interior of the cell), where a realistic description cannot be achieved without the
explicit consideration of diffusion. Here we use our model to study fluctuations in the aggregation process 
induced by small volumes and to provide predictions for the reliability of a seed detection assay.

\section*{Three dimensional model}

\begin{figure}[htb] \centering 
	\includegraphics[width=\columnwidth]{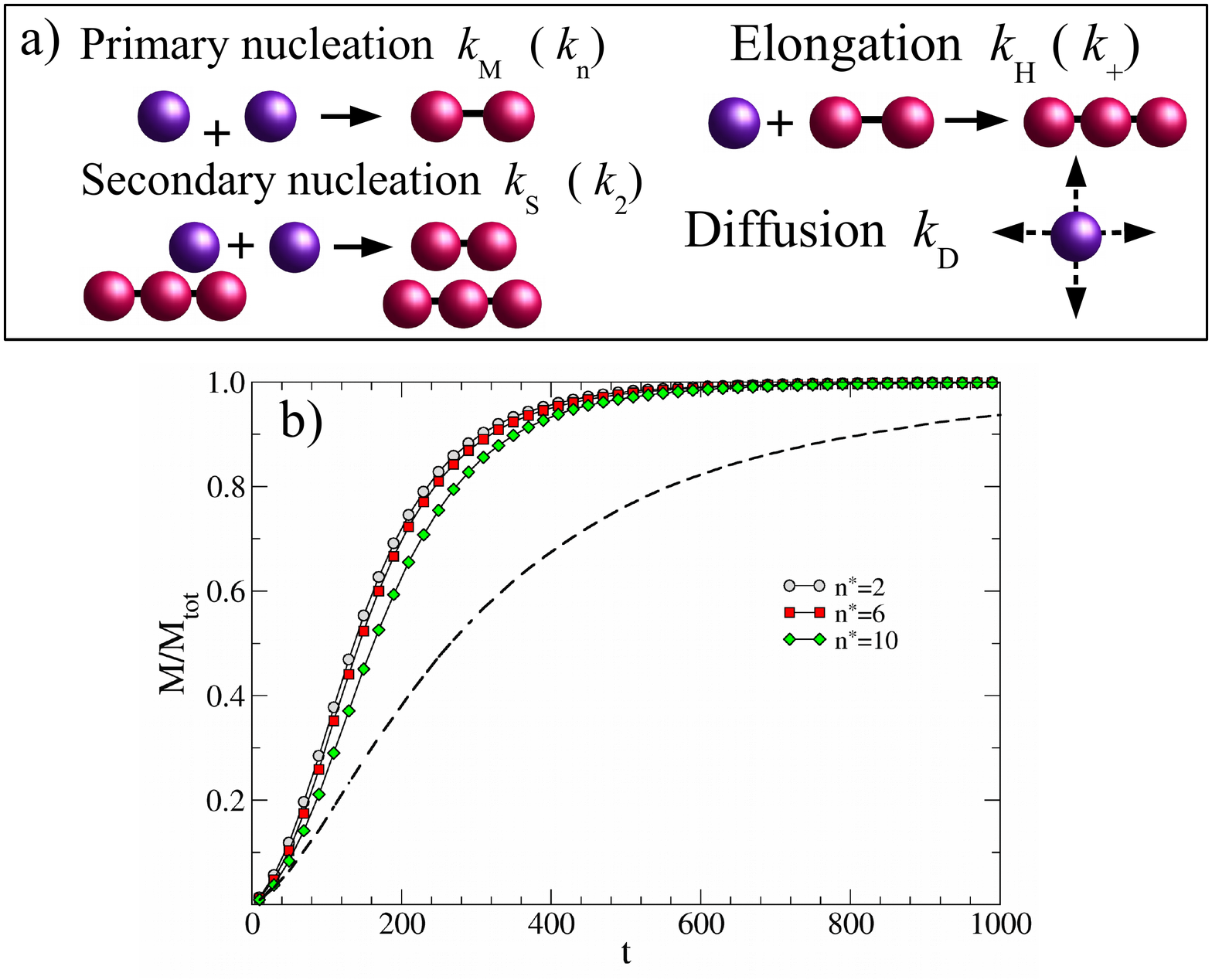}
	\caption{\label{fig:1} 
		a)  Schematic of the protein aggregation model describing the main processes involved:
		primary nucleation occurring with rate $k_{\rm M}$ (and correspondingly $k_{\rm n}$ in mean-field), 
		polymer elongation with rate $k_{\rm H}$ ($k_{\rm +}$ in mean-field), secondary nucleation with rate $k_{\rm S}$ ($k_{\rm 2}$ in mean-field) and
		diffusion with rate $k_{\rm D}$ which is not described by mean-field theory.
		b)  Simulations showing the dependence on the minimal polymer size $n^*$ needed to catalyze secondary nucleation of the
		aggregation curve, the polymer mass fraction $M/M_{\rm tot}$, obtained for
		$k_{\rm M}=4\cdot 10^{-4}$, $k_D=10^{-2}$, $k_{\rm S}=1$, $k_{\rm H}=10^4$ and $\rho=0.16$. The dashed line
		is the curve obtained in the limit $n^* \to \infty$, or equivalently in absence of
		secondary nucleation.}
\end{figure}

Simulations are performed using a variant of the protein aggregation model described in Ref. \cite{budrikis2014} where individual protein molecules sit on 
a three dimensional cubic lattice. The model considers primary nucleation due to monomer-monomer interaction, polymer elongation due to addition of monomers to 
the polymer endpoints and secondary nucleation processes in which the rate of monomer-monomer interaction is enhanced when the process occurs close to a polymer 
(see Fig. \ref{fig:1}a for an illustration)~\cite{Cohen2011c}. In particular, monomers diffuse with rate $k_{\rm D}$ and attach to neighboring monomers with rate 
$k_{\rm M}$ (primary nucleation) but when a monomer is nearest neighbor to a site containing a polymer composed
of at least $n^*$ monomers, then the nucleation rate increases to $k_{\rm S} \gg k_{\rm M}$ (secondary nucleation). We do not consider polymer fragmentation, since 
this term is mostly relevant for samples under strong mechanical action \cite{Cohen2013}, and some of the most important amyloid-forming proteins have been shown to 
exhibit aggregation kinetics dominated by secondary nucleation under 
quiescent conditions~\cite{Cohen2013, Buell2014}. A monomer can attach to a polymer with rate $k_{\rm H}$ if it meets its endpoints. Polymers move collectively by 
reptation with a length-dependent rate $k_{\rm R}/i^2$, where $i$ is the number of monomers in the polymer (see Ref. \cite{Binder1995} p. 89), and locally by 
end rotations, with rate $k_{\rm E}$, and kink moves with rate $k_{\rm K}$ (for a review of lattice polymer models see \cite{Binder1995}). Simulations start 
with a constant number of $N$ monomers in a cubic system of size $L= m_0 L_0$ (with $m_0$ an integer) where $L_0$ is the typical monomer diameter, with periodic 
boundary conditions in all directions. We perform numerical simulations using the Gillespie Monte Carlo algorithm \cite{Gillespie1976} and measure time in units 
of $1/k_{\rm S}$ and rates in units of $k_{\rm S}$. We explore the behavior of the model 
by varying independently both the monomer concentration $\rho\equiv N/m_0^3$ and the number of monomers $N$ at fixed $\rho$, but also
the rate constants. For the simulations results reported in the following, the rates describing polymer motion are chosen
to be $k_{\rm E}=k_{\rm R}=k_{\rm K}=10^{-2}$ which is smaller or equal than the diffusion rate of the monomers $k_{\rm D}$.

As expected, secondary nucleation efficiently decreases the half-time before rapid polymerization. We 
illustrate this by changing the critical polymer size $n^*$ needed to induce secondary nucleation. We observe that the lower $n^*$ the shorter the half-time (see Fig. \ref{fig:1}(b)). Currently, no experimental data exists on the value of $n^*$, 
but it can be expected to be of a similar magnitude as the smallest possible amyloid fibril, defined as the smallest structure for which monomer addition becomes independent of the size of the aggregate and an energetic downhill event. 

\section*{Mean-field theory}

The progress of reactions observed experimentally in bulk systems can be well approximated by a mean-field model \cite{Knowles2009, Cohen2011, Cohen2013}, without fragmentation or depolymerization of polymers. Such models are in contrast to our three dimensional computational model, which describes also monomer diffusion and polymer motion due to reptation, kink motion and end-rotations, which are not treated by mean-field approximation. Despite this, it is still possible to fit polymerization
curves resulting from three dimensional simulation results through mean-field theory with effective diffusion dependent parameters. The fact that both experimental and simulated polymerization curves are described by the same mean-field theory ensures that our model is appropriate to describe experiments. 
In the mean-field model, the evolution of the concentration $f_j$ of polymers of length $j\geq n_c$, where $n_c$ is the nucleation size, is given by~\cite{Cohen2011}
\begin{equation}
\label{eq_polymer_evolution}
\dot{f}_j(t) = k_n m(t)^{n_c} \delta_{j,n_c} + 2m(t)k_{+} f_{j-1}(t)-2m(t)k_{+}f_j(t) +  k_2 m(t)^{n_2} \sum_{i=n_c}^{\infty}i f_i(t) \delta_{j,n_2},
\end{equation}
where dots indicate time derivatives and $m(t)$ is the monomer concentration.
The first term on the right-hand side represents an increase in the concentration of polymers of size $n_c$ due to polymer nucleation by aggregation of $n_c$ monomers with rate constant $k_n$; this is a generalized version of the dimer formation with rate constant $k_{\rm M}$ in the 3d model. The second term represents an increase in the concentration of polymers of size $j$ by attachment of a monomer to a polymer of size $j-1$, with rate constant $k_{+}$. The third term is the corresponding loss of concentration of polymers of size $j$ when they attach to a monomer. These two terms are the mean-field equivalent of the endpoint attachment of monomers to polymers with rate constant $k_{\rm H}$ in the 3d model. The final term represents secondary nucleation, which in the mean-field model is described as an increase in concentration of polymers of size $n_2$ (the secondary nucleus size) occurring at a rate proportional to the mass of polymers and with a rate constant $k_2$.
By conservation of mass, the evolution of the monomer concentration is
\begin{equation}
\label{eq_monomer_evolution}
\dot{m}(t) = - \sum_{i=n_c}^{\infty} i \dot{f}_i(t)
\end{equation}

\begin{figure}[thb] \centering
	\includegraphics[width=\columnwidth]{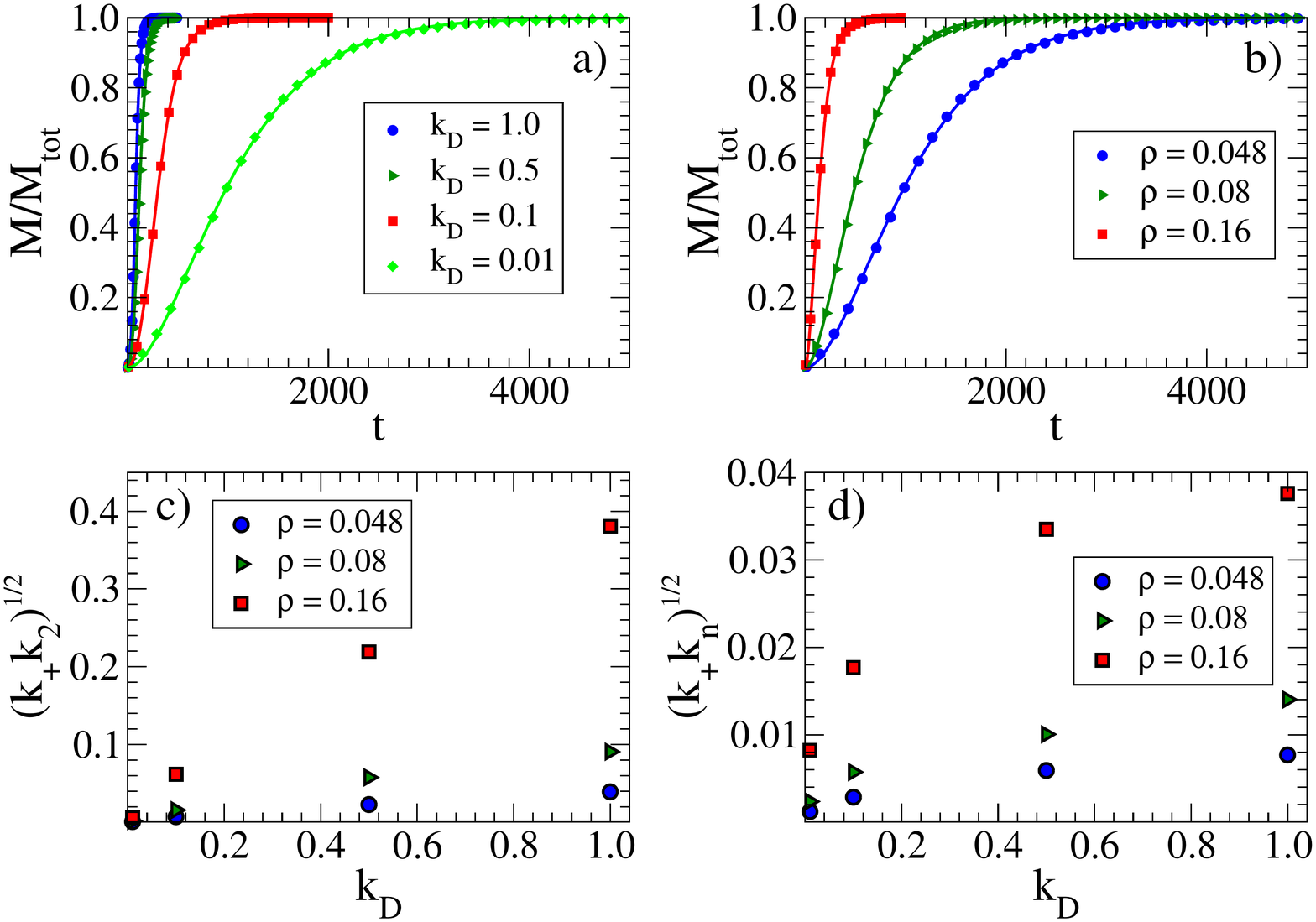} 
	\caption{\label{fig:2} {\bf Protein aggregation depends on monomer diffusion:} 	
		a)  Simulations showing the diffusion dependence of the
		aggregation curve, the polymer mass fraction $M/M_{\rm tot}$, obtained for
		$k_{\rm M}=4 \cdot 10^{-4}$, $k_{\rm S}=k_{\rm H}=1$, $N=2450$ and $\rho=0.048$. The curves are
		well fit by mean-field theory (full lines) with effective parameters that depend
		on $k_{\rm D}$. b) Simulations of the 
		density dependence of the polymer mass fraction for  $N=2450$,
		$k_{\rm M}=4 \cdot 10^{-4}$, $k_{\rm H}=1$ and $k_{\rm D}=10^{-2}$. Fits by mean-field theory are plotted
		as lines with effective parameters reported in the panels c) and d). Time is measured in units of $1/k_{\rm S}$.
		c)-d):The effective mean-field parameters $\sqrt{k_+k_2}$ and $\sqrt{k_+k_n}$ obtained by fitting 
		simulations performed for $k_{\rm M}=4\cdot10^{-4}$, $k_{\rm S}=k_{\rm H}=1$ as a function
		of the concentration $\rho$ and the diffusion rate $k_{\rm D}$.
		}
\end{figure}

The evolution of the number concentration $P(t)=\sum_{j\geq n_c} f_j(t)$ and mass concentration $M(t)=\sum_{j\geq n_c} j f_j(t)$ can be found by summing over $j$ in \eqref{eq_polymer_evolution}. After some algebra, one obtains
\begin{gather}
\dot{P}(t) = k_2 M(t) m(t)^{n_2} + k_n m(t)^{n_c},\\
\dot{M}(t) = 2 k_{+} m(t) P(t) + n_2 k_2 m(t)^{n_2} + n_c k_n m(t)^{n_c},
\end{gather}
Analytical approximation~\cite{Knowles2009, Cohen2011, Cohen2013} of the system of equations gives a solution that depends on two parameters, 
$\lambda=\sqrt{2k_+k_nm(0)^{n_c}}$ and $\kappa=\sqrt{2k_+k_2m(0)^{n_2+1}}$. We fit our data with the form given in Eq.~1 of Ref.~\cite{Cohen2013},  using a least squares method. Each curve is fitted independently.
Diffusion plays an important role in the aggregation progress, shifting the aggregation curves as shown in Fig. \ref{fig:2}a and Fig. \ref{fig:2}b. 
For a considerable parameter range, however, the time evolution of the fractional polymer mass can be fitted by mean-field theory (lines in Fig. \ref{fig:2}a and Fig. \ref{fig:2}b) with effective parameters that now depend on the diffusion rate $k_{\rm D}$ (see Fig. \ref{fig:2}c and \ref{fig:2}d).  Similarly, mean-field theory describes the density dependence of the aggregation curves as shown in Fig. \ref{fig:2}b.

\section*{Fluctuations in protein aggregation}
\begin{figure}[htb]
	\centering
	\includegraphics[width=5.2in]{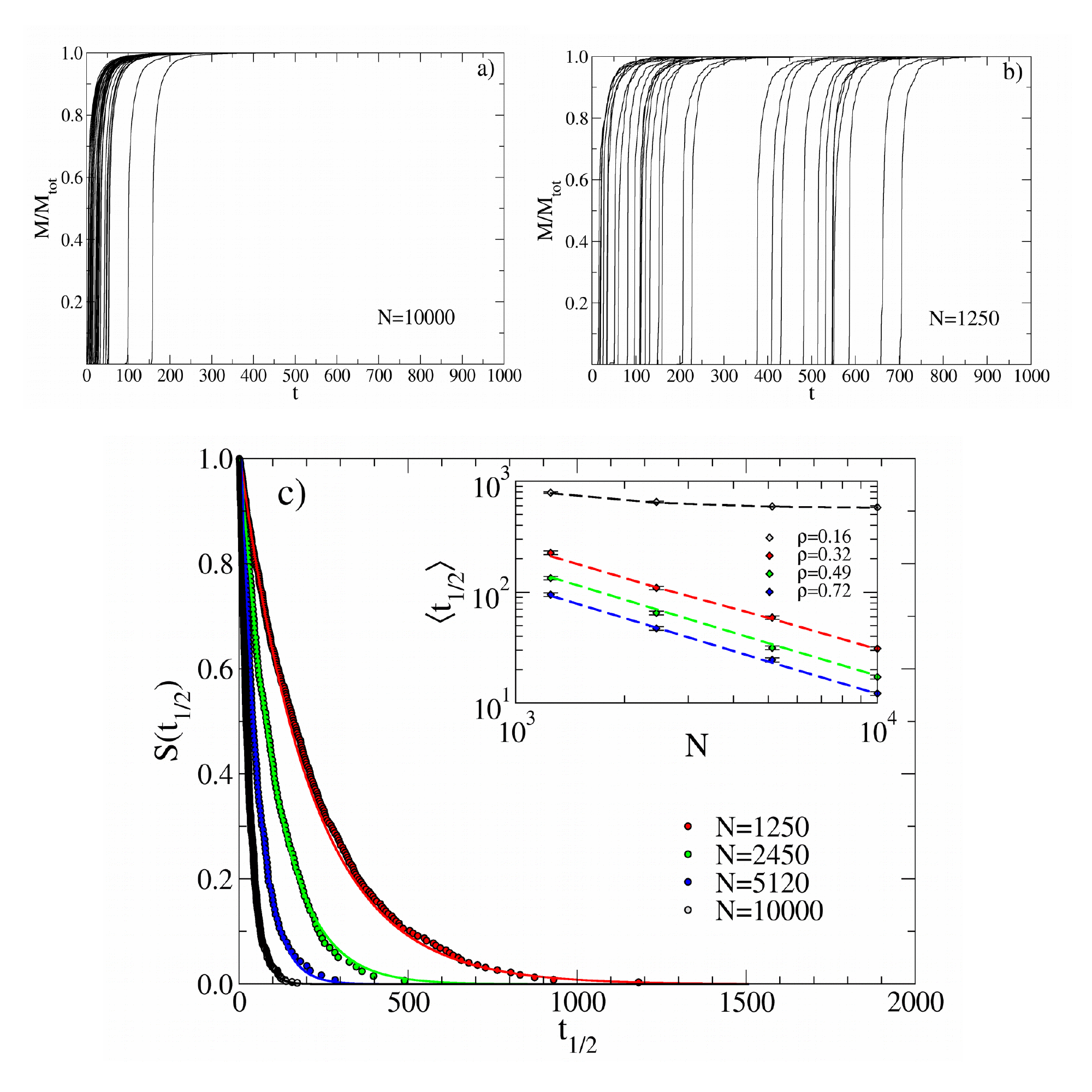} 
	\caption{\label{fig:3} {Half-time sample-to-sample fluctuations are due to
		extreme value statistics}. Different replicates of the simulations display 
		wide fluctuations in half-times especially for small numbers of monomers. 
		a) Simulation results obtained for $N=10000$ monomers at $\rho=0.32$,
		$k_{\rm M}=4 \cdot 10^{-6}$, $k_{\rm S}=1$, $k_{\rm H}=10^4$, 
		$k_{\rm D}=10^{-2}$. The graph shows that the half-time $t_{1/2}$ is very close
		to the nucleation time $t_0$ at which the curves depart from zero.
		b) Same as panel a) but with $N=1250$. c) The complementary cumulative distributions
		of half-times obtained from simulations for different values of $N$ are 
		in agreement with the theory described in the text.
		The inset shows that the average half-times for different concentrations $\rho$ as a function
		of  $N$. The general trend is in agreement	with experiments \cite{Knowles2011}. Time is measured in units of $1/k_{\rm S}$.} 
\end{figure}

Having confirmed that our computational model faithfully reproduces polymerization kinetics in macroscopic samples, we now turn to the main focus of the paper, 
the study of sample-to-sample fluctuations in small volumes, a feature that can not be studied with mean-field kinetics. When the sample volume is reduced, 
we observe increasing fluctuations in the aggregation kinetics as shown in Figs. \ref{fig:3}a and \ref{fig:3}b. These results are summarized in Fig \ref{fig:3}c 
showing the complementary cumulative distributions of half-times for different monomer numbers $N$ and constant monomer concentration 
\begin{equation}
S(t_{1/2}) \equiv \int_{t_{1/2}}^\infty P(x)dx
\end{equation}
where $P(x)$ is the probability density function and $t_{1/2}$ is defined as the half-time of the polymerization curve (i.e. the time at which $M/M_0=1/2$).  

\begin{figure}[h]
	\begin{center}
		
		\includegraphics[width=4.5in]{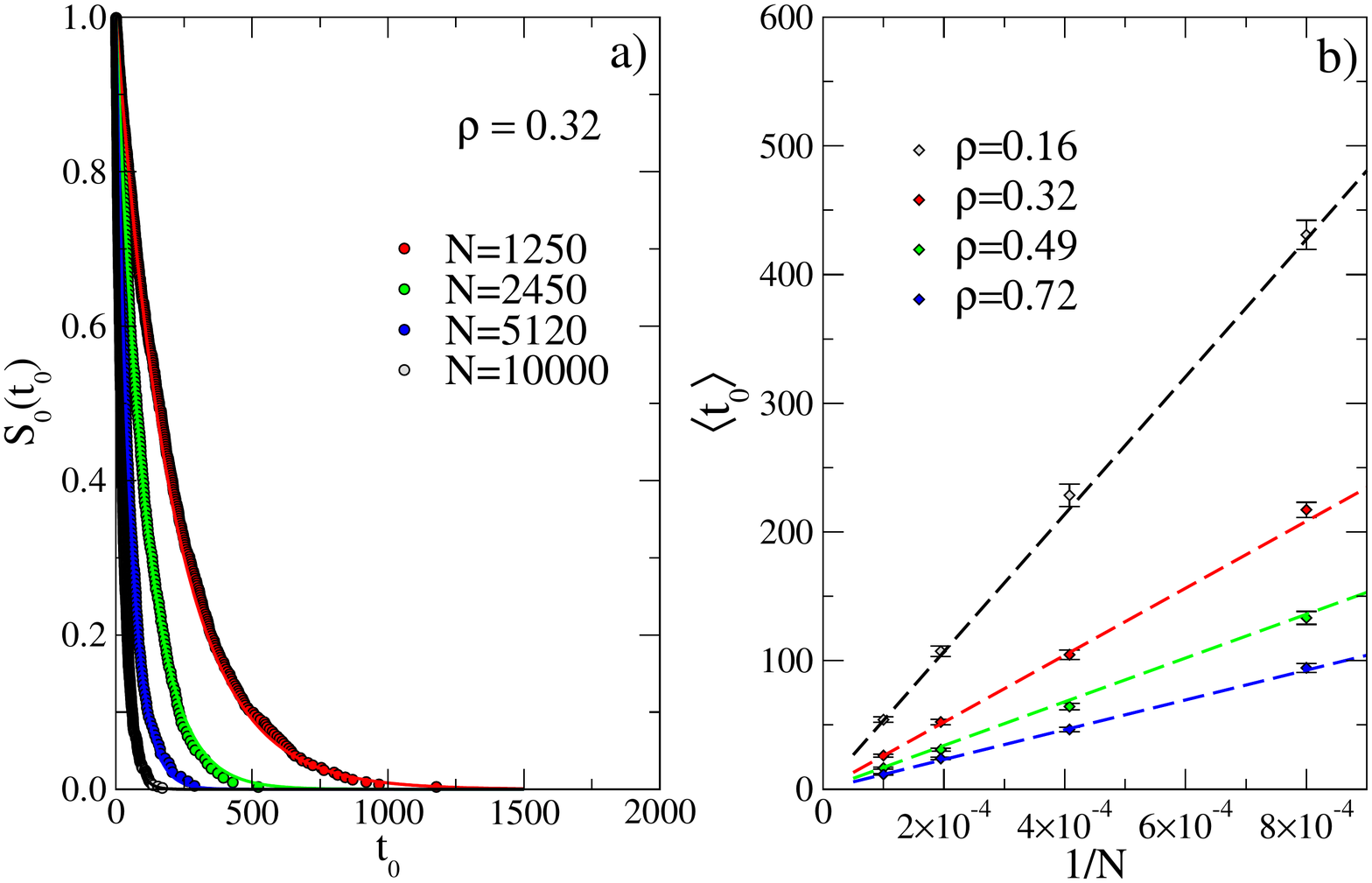}
		\caption{ a) The complementary cumulative distribution functions $S(t_0)$ for four different monomer numbers $N$ and density $\rho=0.32$. 
			The symbols correspond to the numerical simulations while lines correspond to the theoretical predictions obtained from Eq. (6). 
			b) The average time $\langle t_0 \rangle$ as a function of $1/N$ for four different number densities. The theoretical predictions 
			(dashed lines) are obtained  from Eq.(13). Here $k_D=10^{-2}$ and $k_M=4 \cdot 10^{-6}$
		}
		\label{figS0_32}
	\end{center}
\end{figure}

\begin{figure}
	\begin{center}
		
		\includegraphics[width=4.5in]{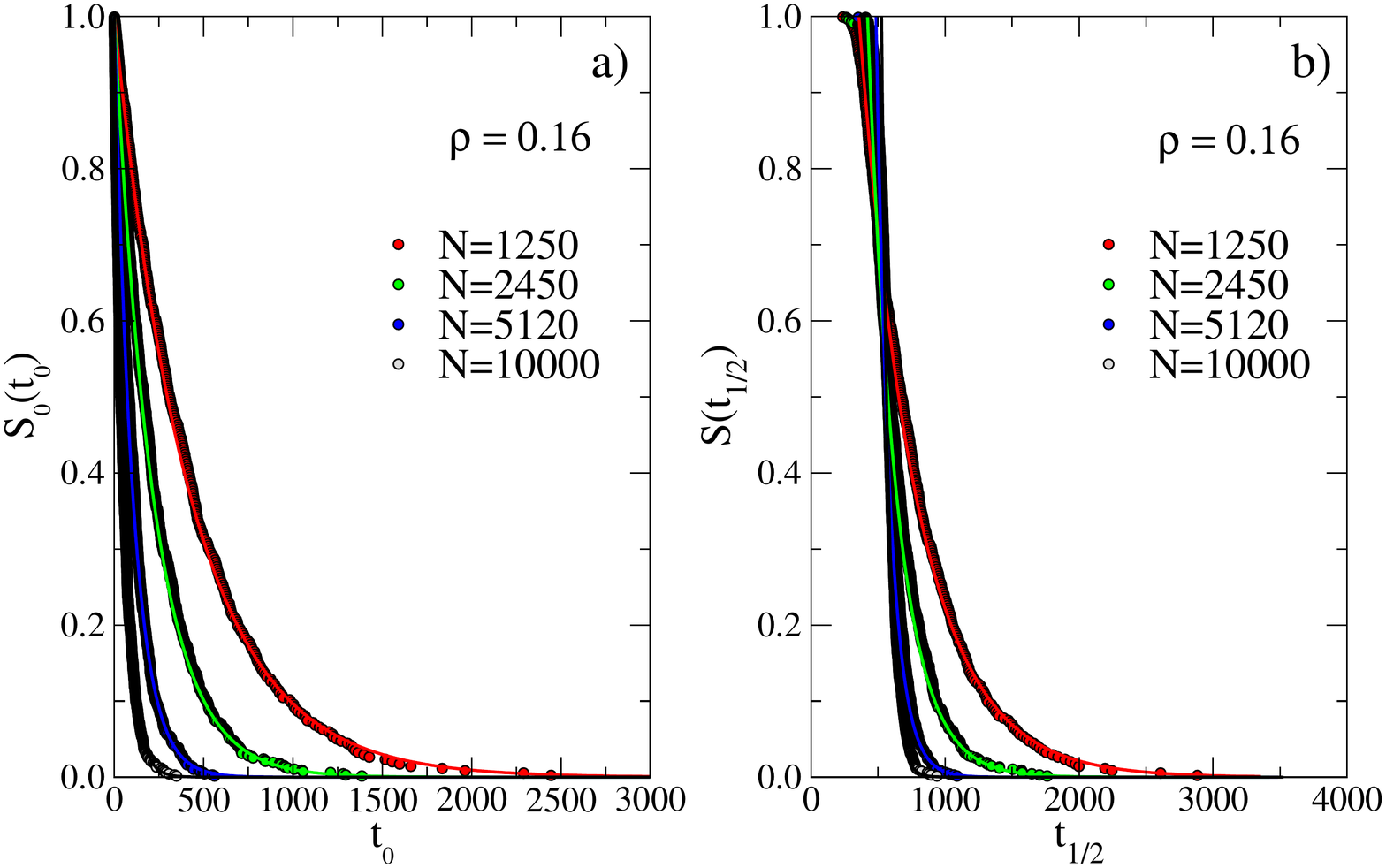}
		\caption{The complementary cumulative distribution functions $S(t_0)$ (a) and $S(t_{1/2})$ (b) for four different monomer numbers $N$ and 
			density $\rho=0.16$. 
			The symbols correspond to the numerical simulations while
			the lines represent the theoretical predictions obtained from Eq. (6) and Eq. (7).
			Here $k_D=10^{-2}$ and $k_M=4 \cdot 10^{-6}$.
		}
		\label{figS0_16}
	\end{center}
\end{figure}

\begin{figure}
	\begin{center}
		
		\includegraphics[width=4.5in]{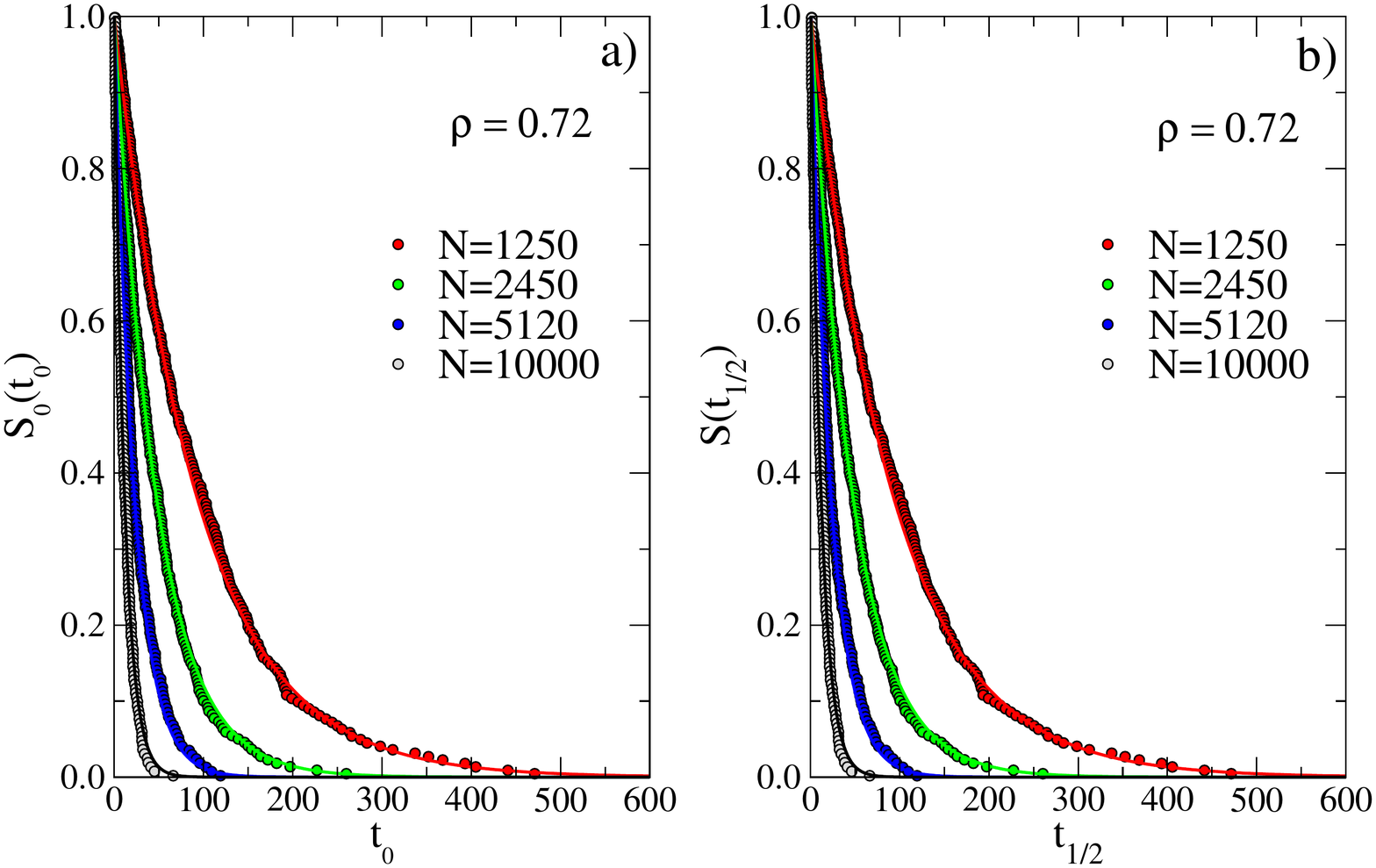}
		\caption{The complementary cumulative distribution functions $S(t_0)$ (a) and $S(t_{1/2})$ (b) for four different monomer numbers $N$ 
		and density $\rho=0.72$. 
			The symbols correspond to the numerical simulations while
			he lines represent the theoretical predictions obtained from Eq.(6) and Eq.(7).
			Here $k_D=10^{-2}$ and $k_M=4 \cdot 10^{-6}$. }
		\label{figS0_72}
	\end{center}
\end{figure}

\begin{figure}
	\begin{center}
		
		\includegraphics[width=4.5in]{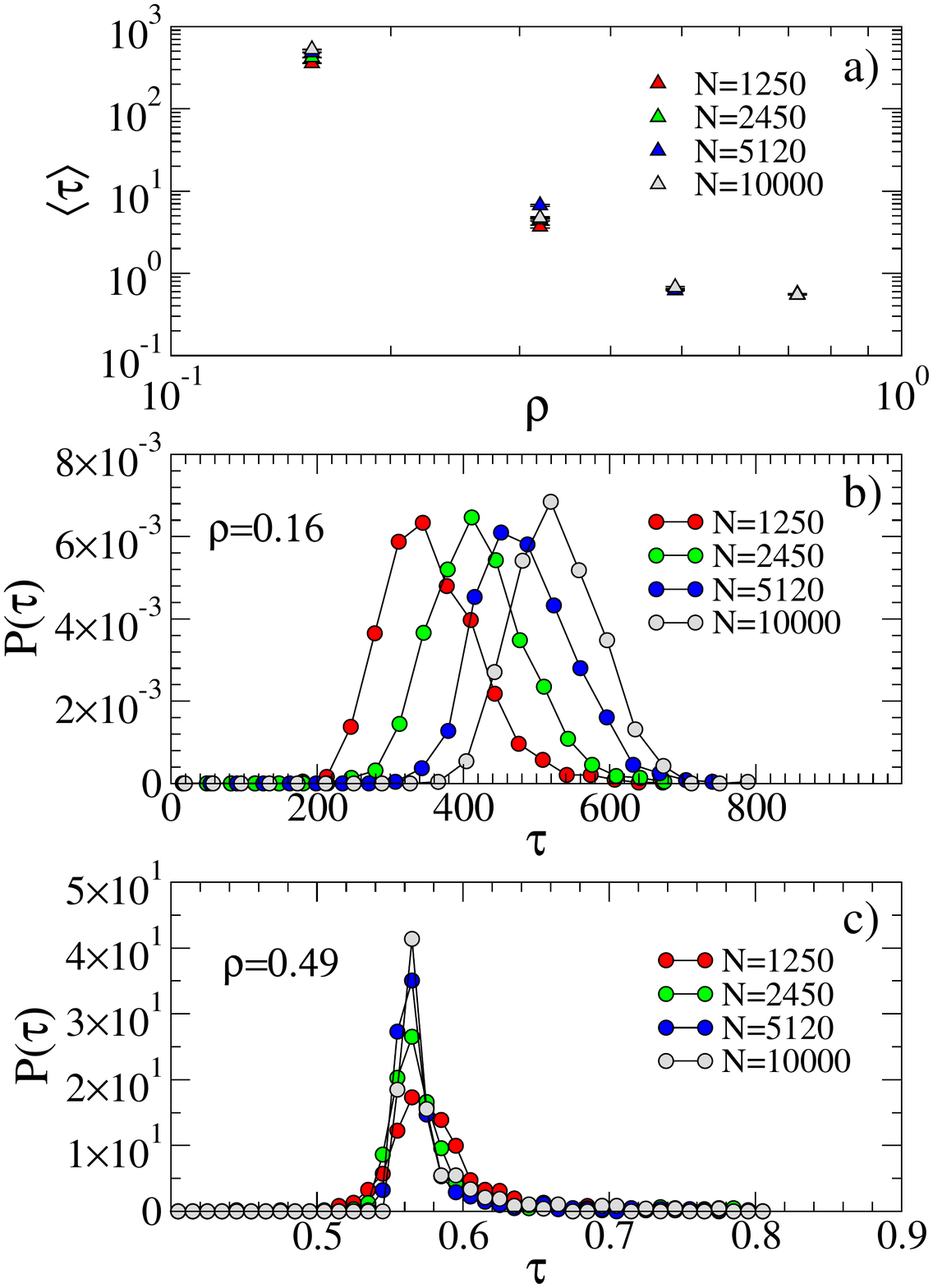}
		\caption{ 
			a)  The mean delay time $\langle \tau\rangle$, obtained from numerical simulations, as a function of the number density $\rho$ 
			for four different monomer numbers $N$. For any $N$ and $\rho$ 
			the averages are calculated over the different numerical simulations outcomes. The distributions of delay times $\tau$ as a function of the number of monomers $N$
			for b) $\rho=0.16$ and c) $\rho=0.49$. Here $k_D=10^{-2}$ and $k_M=4 \cdot 10^{-6}$.}
		\label{figSalpha}
	\end{center}
\end{figure}

The steepness of the aggregation curves in Figs. \ref{fig:3}a and \ref{fig:3}b 
suggests that, for $k_S\gg k_M$, fluctuations are mostly ruled by the time of the first primary nucleation event,
$t_0$, whose complementary cumulative distribution $S_0(t_0)$ can be estimated analytically as 
\begin{equation}
S_0(t_0)=e^{-f_{M}k_MNt_0}
\label{eq:S0}
\end{equation}
where $f_{M}$ is the average number of possible primary nucleation events per unit monomer. We estimate that
$f_M =3\rho$ using a Poisson approximation, as we show in details in the following
section. Note that Eq. \ref{eq:S0} displays a size dependence that is reminiscent of extreme value statistics   $S_0(x,N) = \exp(-N F(x))$ where $F(x)$ is a function that does not depend on $N$
\cite{gumbel,weibull39}. If $k_S \gg  k_M$, the half-time $t_{1/2}$ differs from the nucleation
time by a weakly fluctuating time $\tau$. This comes from the  observation that, once the first primary nucleation event has happened, the polymerization follows rapidly, 
thanks to fast growth and secondary nucleation. 
This yields a weakly fluctuating delay $\tau(N,\rho)=t_{1/2}-t_0$, which in general  depends on the number density $\rho$ and on the number of monomers $N$. The distribution and average values of $\tau$ are reported in Fig. \ref{figSalpha}. The average value of $\tau$ decreases with $\rho$ and
displays only a smaller dependence on $N$. The distribution of $\tau$ is always peaked around its
average but while at small values of $\rho$ the peak shifts with $N$ while the standard
deviation remains constant, for higher values of $\rho$
only the standard deviation depends on $N$ and the peak position does not change.
Since the fluctuations in $\tau$ are much smaller than the fluctuations of $t_0$ we can safely assume that $t_{1/2}\simeq t_0+\langle \tau\rangle$ for $t_0\geq 0$, so
the complementary cumulative distribution takes the form 
\begin{equation}
S(t_{1/2}) =\left\{
\begin{array}{ll}
1 & t_{1/2}\leq \langle \tau\rangle\\
S_0(t_{1/2}-\langle\tau\rangle)  & t_{1/2}> \langle \tau \rangle.
\end{array}
\right.
\label{eq:Slag}
\end{equation}  
The prediction of Eqs. \ref{eq:S0} and \ref{eq:Slag} are in agreement with numerical 
simulations results for $S_0(t_0)$ and $S(t_{1/2})$, respectively (see Figs. \ref{fig:3}c and \ref{figS0_32}). In particular, the behavior of $S_0(t_0)$  is obtained without any fitting parameters, while $S(t_{1/2})$ only needs the estimate of the single parameter $\langle \tau \rangle$  (additional comparisons for different  values of $\rho$ are reported in Figs. \ref{figS0_16} and \ref{figS0_72}). 
The corresponding average values of $\langle t_{1/2}\rangle$ are shown in the inset of Fig. \ref{fig:3}c as a function of $N$ (see also Fig. \ref{figS0_32}b).

\section*{Theoretical derivation of the half-time distribution}

\begin{figure}
	\begin{center}
		
		\includegraphics[width=4.5in]{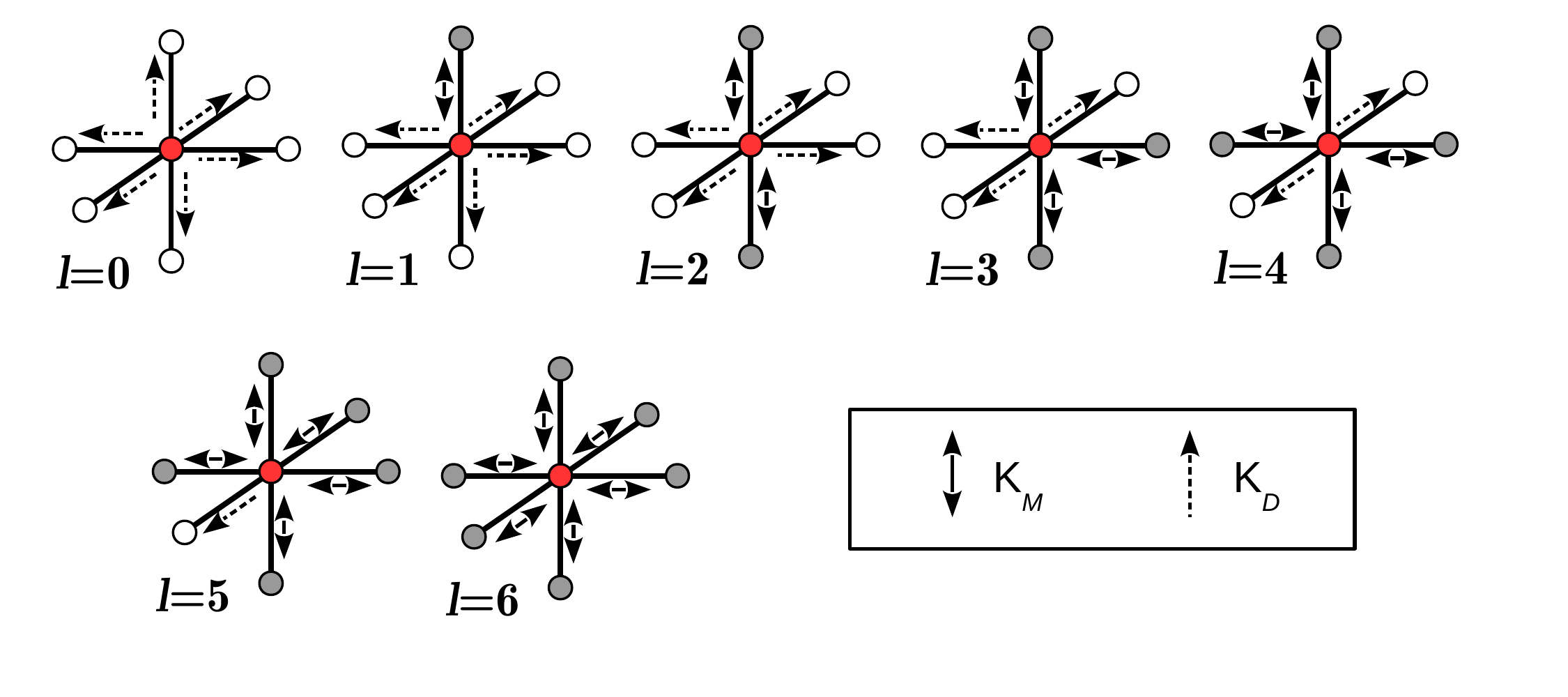}
		\caption{A schematic representation of the possibilities of diffusion (dashed arrow) and aggregation (double arrow) for a monomer (red circle) 
			placed in the center of cubic lattice unit cell. 
			The monomer partners for the dimerization (from $l=1$ to $l=6$) are colored in grey while the empty sites are represented by white circles. 
			The aggregation and diffusion rate are, respectively, $k_M$ amd $k_D$.
		}
		\label{fig:lattice}
	\end{center}
\end{figure}

\begin{figure}
	\begin{center}
		
		\includegraphics[width=\columnwidth]{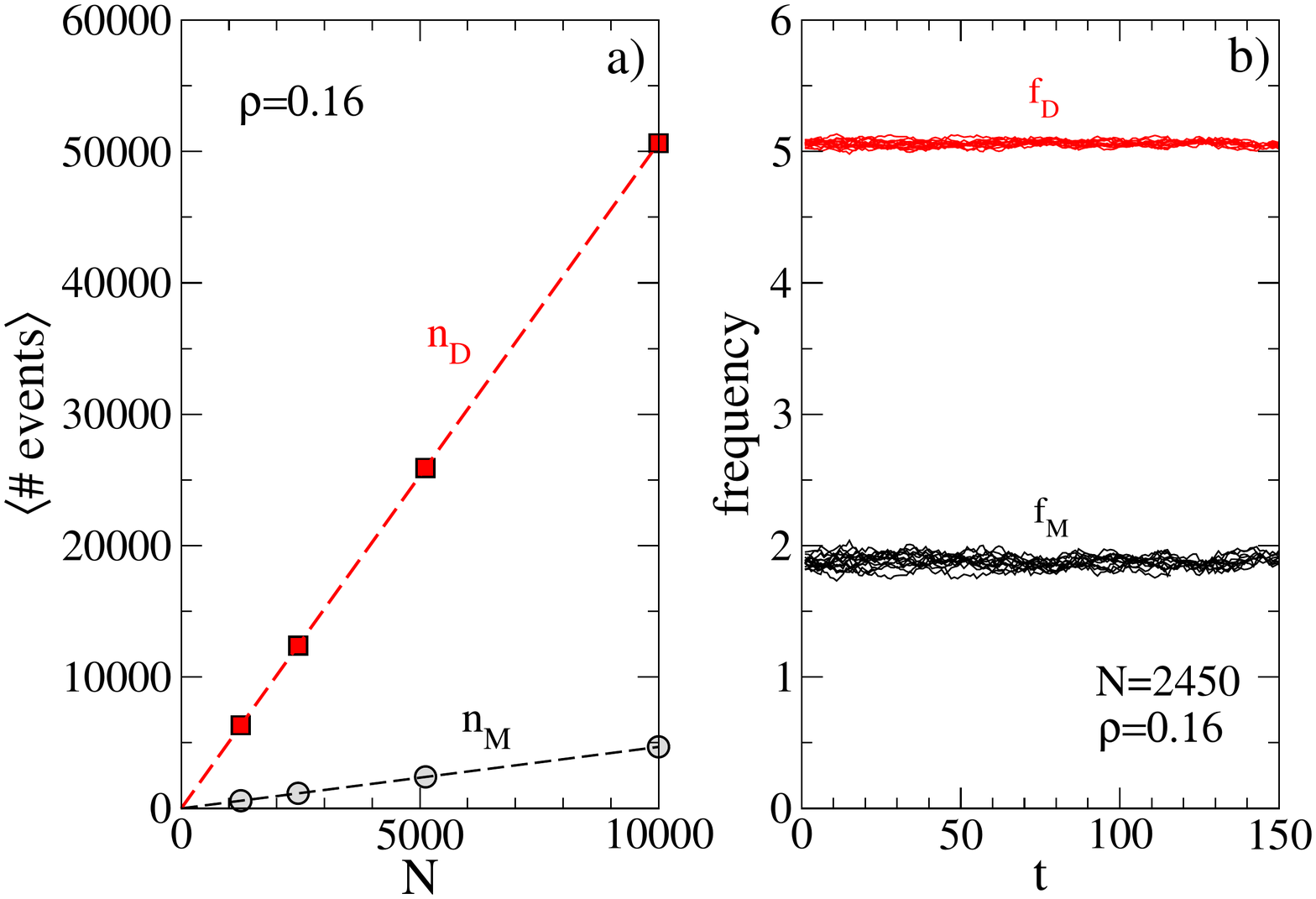}
		\caption{ a) The number of possible primary nucleation and diffusion events 
			($n_M$ and $n_D$, respectively) are a linear function of the number of monomers $N$. 
			b) The corresponding frequencies $f_M$ and $f_D$ fluctuate very little in time. 
			Here $k_D=10^{-2}$, $k_M=4 \cdot 10^{-6}$,  $N=2450$ and $\rho=0.16$.
		}
		\label{fig:ergodic}
	\end{center}
\end{figure}

In this section, we provide a detailed derivation of Eqs. \ref{eq:S0} and \ref{eq:Slag} in the limit
of relatively large diffusion when the system is well mixed. To this end,
we consider a cubic lattice composed by $m_0^3$ nodes, in which $N$ monomers are placed randomly at time $t=0$. 
As illustrated in Fig. \ref{fig:lattice}, each monomer $i$ sits near $l^{(i)}$ neighboring monomers and $6-l^{(i)}$ neighboring empty sites, where $l^{(i)}$ is in general a fluctuating time dependent quantity. 
In the model, each monomer $i$ can either diffuse into one of the $6-l^{(i)}$  empty sites or form a dimer with one of the 
$l^{(i)}$ neighboring monomers.  Therefore, at any given time $t$ the number of possible 
diffusion events in the system is $n_D(t) = \sum_i (6-l^{(i)})$ and the number of possible aggregation
events is $n_M(t) = 1/2\sum_i l^{(i)}$, where the factor $1/2$ is needed to correct for the double counting
of the number of monomer pairs. We can compute the time of first aggregation of $N$ monomers
using Poisson statistics, considering for simplicity the case in which
the number of possible aggregation events $n_M$ would not depend on time. In this case, 
the probability of having an aggregation event within  $\Delta t$ is $n_{M}k_M\Delta t$. 
We can then divide the time interval $t_0$ in $n$ elementary time subintervals 
$\Delta t=\frac{t_0}{n}$, the rate of aggregation event at $t_0$, i.e. the probability per unit time to have the first dimer formed after  a time interval $t_0$ has elapsed, is given by the following expression: 
\begin{equation}
\tilde{P}_0(t_0)=\lim_{n\to\infty}\left(1-n_{M}k_M\frac{t_0}{n}\right)^{n-1}n_{M}k_M=n_Mk_Me^{-n_{M}k_Mt_0}.
\label{P0_t0}
\end{equation}
As stressed previously, $n_M$ and $n_D$ are in principle fluctuating quantities and therefore Eq.\ref{P0_t0} is not valid. Yet, as shown in
Fig. \ref{fig:ergodic}: $n_M$ and $n_D$ are are both $i)$ stationary, $ii)$ ergodic, $iii)$ weakly fluctuating and $iv)$ linearly  dependent on $N$, on average. Hence, the probability  $\tilde{P}_0(t_0)$ for a monomer to form a dimer at $t_0$ can reasonably be approximated by its ensemble average
\begin{equation}
P_0(t_0)\simeq \langle \tilde{P}_0(t_0)\rangle \simeq \langle n_{M}\rangle k_M e^{-f_{M}k_MNt_0}
\label{p_real_norm}
\end{equation}
where we have replaced $n_M$ by its average value $\langle n_{M}\rangle$ and defined
$f_{M}\equiv \frac{\langle n_{M}\rangle}{N}$.
From Eq.(\ref{p_real_norm}), we easily obtain the complementary cumulative distribution function 
\begin{equation}
S_0(t_0)=\int_{t_0}^{\infty} d\tau P_0(\tau)=e^{-f_{M}k_MNt_0},
\label{S0_real}
\end{equation}
recovering Eq. \ref{eq:S0}. 

To conclude our calculation, we still need to evaluate $f_M$. To this end, we perform a discrete enumeration of the possible configurations of a single monomer, in the spirit of cluster expansions for percolation models. In particular, the six relevant configurations for a single monomer in contact with other monomers are reported in Fig. \ref{fig:lattice}. The weight $p_{l}$  of a configuration in which a monomer has $l$  occupied neighbors is assumed to be given by the binomial distribution
\begin{equation}
p_{l} = \frac{6!}{l!(6-l)!}\rho^l(1-\rho)^{6-l}
\label{binomial}.
\end{equation}
This single particle picture suggests that the average number of primary nucleation events per monomer $f_{M}$,  
corresponds to the average number of nearest neighbors $\langle l \rangle$, divided by a factor 2 since any nucleation event encompasses 2 particles. With a similar reasoning, we estimate $f_{D}=6 - \langle l \rangle$, 
i.e. the average number of empty directions. Then, from the binomial distribution (\ref{binomial}) we get  $f_{M}=\frac{6\rho}{2}$ and $f_{D}=6(1-\rho)$. Using these values in Eq. \ref{eq:S0} and Eq.(\ref{eq:Slag}), 
we obtain agreement with numerical simulations as illustrated in Figs. \ref{figS0_16}
and \ref{figS0_72} (panels (a) and (b) respectively) for different values of $\rho$.


Finally, we calculate the averages of the first aggregation time and the half-time as 
$\langle t_0\rangle =\int_{0}^\infty dt_0\,t_0P_0(t_0)$ and $\langle t_{1/2}\rangle =\int_{0}^\infty dt_{1/2}\,t_{1/2}P(t_{1/2})$,  where $P(t_{1/2})=-\frac{dS(t_{1/2})}{dt_{1/2}}$.  The expression for $\langle t_0\rangle$ is given by
\begin{equation}
\langle t_0\rangle= \frac{1}{3\rho k_M N}.
\label{mean_t0}
\end{equation}
In Fig. \ref{figS0_32}b  we show the perfect agreement of the theoretical estimate given by Eq.(\ref{mean_t0}) with the numerical values of the  average time for the first primary nucleation event as a function of $1/N$, for several densities. Notice that no fitting parameters are involved. 
The average time of $t_{1/2}$ follows from 
$\langle t_{1/2}\rangle=\langle t_0\rangle+\langle \tau\rangle$: the inset of Fig. \ref{fig:3}c confirms the agreement between the theoretical 
estimates  (dashed lines) and the numerical values (symbols).

\section*{Statistical analysis of seed detection tests}

While the fluctuations we observe are intrinsic to the random nature of nucleation events, the ones usually encountered in bulk experiments are likely due to contamination or differences in initial conditions \cite{Giehm2010, Uversky2001, Grey2015}. In those bulk systems ($\mu$l and larger), the number of protein molecules involved in the aggregation process is extremely large, even at low concentrations, so that we can exclude intrinsic kinetic fluctuations. For instance, a volume of 100$\mu$l at a concentration of 1$\mu$M still contains $10^{14}$ monomers, leading to a large number (hundreds to thousands) of nucleation events per second for a realistic value of the nucleation rate~\cite{Cohen2013, Buell2014a}. However, if the relevant volumes are made small enough (pico- to nanolitres), the stochastic nature of primary nucleation can be directly observed. This has been exploited by aggregation experiments performed inside single microdroplets, where individual nucleation events could be observed, due to their amplification by secondary processes~\cite{Knowles2011}. In these experiments, the average half-time is found to scale with volume in a similar manner to what is shown in the inset of Fig. \ref{fig:3}, thus in agreement with our simulations.  

\begin{figure}[thb]
	\centering
	\includegraphics[width=\columnwidth]{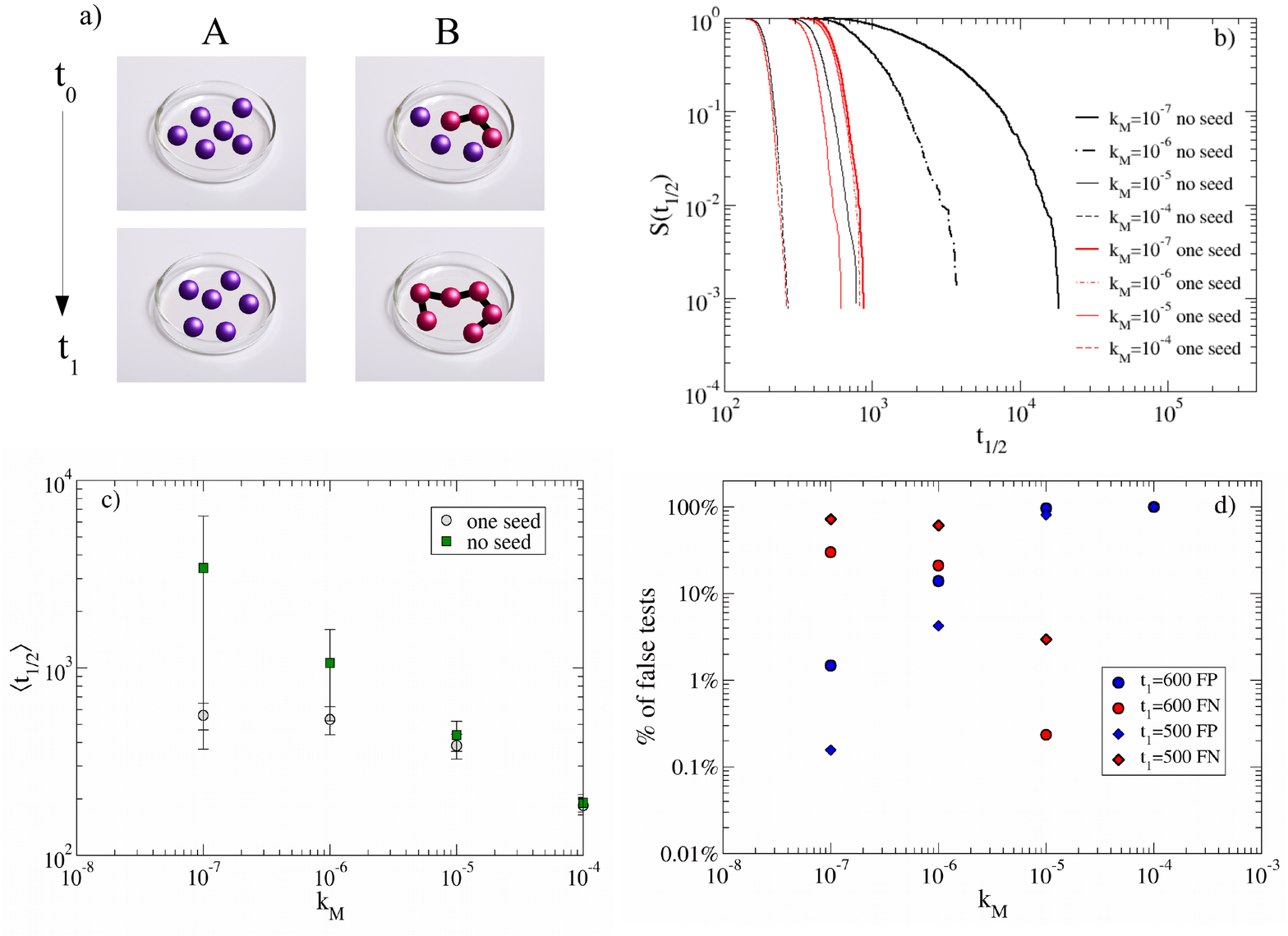} 
	\caption{\label{fig:6} {Intrinsic fluctuations rule errors in the detection of protein aggregation
			prone samples}. a) A test to detect seeds for protein aggregation is based on
		small volume sampling of protein solutions at time $t_0=0$ and on the assumption that
		only seeded samples (e. g. sample B) would form aggregate at time $t_1$. b) Simulations
		allow to compute the distribution of half-times for samples with and without a
		seed as a function of the rate of primary nucleation $k_M$. Data are obtained sampling over
		$n=1200$ independent realization. c) Average half-time ($\pm$ standard deviation) 
		as a function of the rate of primary nucleation $k_M$. d) Fraction of false
		positives (FP) and false negatives (FN) for testing times $t_1=600$ and $t_1=500$. 
		Time is measured in units of $1/k_{\rm S}$, $N=1250$.
	} 
\end{figure}

We are now in a position to use our model to design a test {\it in silico} to detect the presence of pre-formed polymers, that act as seeds and nucleation sites for 
the secondary nucleation process, and that are thus amplified. As illustrated in Fig. \ref{fig:6}a, the test considers a set of small volume samples containing protein solutions with a given concentration at time $t_0=0$. 
The aim of the test is to detect the samples containing at least one seed (case B in Fig. \ref{fig:6}a). 
In an ideal experiment, the size of the microdroplets would be adjusted such that most droplets contain no seeds, some contain one seed, and the proportion of droplets 
containing more than one seed is negligible. In practice, these conditions can be easily adjusted experimentally by progressively reducing droplet volumes until only a 
small fraction of them display aggregates. After a fixed time $t_1 \sim \langle \tau \rangle$, one can observe which samples contain macroscopic, detectable amounts of 
aggregates, enabling exact quantification of the number of aggregates present in the initial sample. Ideally the test should be positive only when at least one seed was 
initially present, but given the large fluctuations intrinsic to the nucleation processes we demonstrated above, as well as the competition with \textit{de novo} nucleation, 
there is a chance for false tests. In particular, a false positive test occurs when an unseeded sample is found to contain aggregates, while a false negative test corresponds 
to the case in which a seeded sample does not produce detectable amounts of aggregates within the time scale of the experiment. 

In Fig. \ref{fig:6}b we report the complementary cumulative distribution of aggregation half-times $t_{1/2}$ as a function of the primary nucleation rate $k_{\rm M}$ for 
samples with or without seeds, in this case a single pre-formed trimer. For small values of $k_{\rm M}$, seeded and unseeded samples yield distinct results, as also 
illustrated by the average half-times reported in Fig. \ref{fig:6}c. As the value of $k_{\rm M}$ increases, however, the distributions become closer in the two cases. 
In Fig. \ref{fig:6}d, we quantify the fraction of false positive and false negative tests for two different testing times (e.g. $t_1=500$ and $t_1=600$). As expected, 
for large $k_{\rm M}$ errors are very likely and the test would not be reliable. For intermediate values of $k_{\rm M}$, one can try to adjust $t_1$ to reduce possible 
errors with a caveat: decreasing $t_1$ reduces false positive errors, but at the same time increases false negatives (Fig. \ref{fig:6}d). It is, however, possible to 
optimize $t_1$ so that both types of errors are minimized. In an experimental realisation of such a setup, the most important system parameter that needs to be optimized 
for any given protein is the ratio of secondary nucleation rate to primary nucleation rate. Due to potentially different dependencies of these two rates on the monomer 
concentration~\cite{Meisl2014}, pH~\cite{Buell2014} and potentially other factors, such as temperature, salt concentration etc., it is possible to fine tune this ratio 
and adjust it to a value that allows for an easy discrimination between droplets that do contain a seed aggregate and those that do not.

\section*{Conclusions}

In conclusion, we study protein polymerization in a three dimensional computational model and elucidate the role of protein diffusion in the polymerization process.  Most theoretical studies of protein aggregation neglect completely the role of diffusion and any other spatial effect.
When the polymer diffusion and elongation rates are large enough we recover the standard polymerization curves that can also be obtained from mean-field analytical 
treatments and that can be used to fit, for example, kinetic data of amyloid $\beta$ aggregation~\cite{Cohen2013}. It would be interesting to explore if for small diffusion rates
and small densities mean-field kinetics would eventually fail to describe the results, but this is a challenging computational task. 
At low densities, diffusion could play an important since a critical timescale would be set by the time needed by two  monomers to meet before aggregating. This time-scale can be estimated considering
the time for a monomer to cover a distance $x_D \sim \rho^{-1/3}$, yielding 
$t_D\sim x_D^2/D \sim D^{-1} \rho^{-2/3}$. This timescale is not relevant for our simulations since
at the relatively high densities we study a considerable fraction of monomers are close
to at least another monomer (see $n_M$ in Fig. \ref{fig:ergodic}a). Consequently, the
distribution of the first aggregation time does not depend on the diffusion rate $k_D$
(see Eq. \ref{eq:S0}). The half-time distribution, however, depends on diffusion even
in this regime (Fig. \ref{fig:2}). 

Our simulations show intrinsic sample-to-sample fluctuations that become very large in the limit of small volumes and low aggregation rates. We show that the corresponding half-time distributions 
are described by Poisson statistics and display size dependence. As a consequence of this, the average half-times scale as the inverse of the sample volume, in agreement with insulin aggregation experiments performed in microdroplets \cite{Knowles2011} and with  calculations based on a master-equation approach \cite{Szavits-Nossan2014}. We use this result to design and validate {\it in silico} a pre-clinical screening test  based on a subdivision of the macroscopic sample volume that will ultimately allow the determination of the exact number of aggregates that was initially present. 
This is the first step to develop microarray-based {\it in vitro} tests for early diagnosis of neurodegenerative diseases.

%
%

\section*{Acknowledgements:}
GC,ZB, ATand SZ are supported by ERC Advanced Grant 291002 SIZEFFECTS. CAMLP thanks the visiting professor program of Aalto University where part of this work was completed.  SZ acknowledges support from the Academy of Finland FiDiPro program, project 13282993. AKB thanks Magdalene College, Cambridge and the Leverhulme Trust for support. 




%

%

\clearpage


%
%

\end{document}